\documentclass[10pt]{article}
\usepackage[letterpaper]{geometry}
\usepackage{hicss}
\usepackage{times}
\usepackage[none]{hyphenat}
\usepackage{url}
\usepackage{latexsym}
\usepackage{indentfirst}
\usepackage{graphicx}
\graphicspath{{images/}}

\usepackage{subfigure}
\usepackage{url}
\usepackage{xcolor}

\definecolor{MyP}{HTML}{005493}
\definecolor{MyC}{HTML}{941001}
\definecolor{MyD}{HTML}{FF9402}
\definecolor{MyI}{HTML}{D784FF}
\definecolor{MyT}{HTML}{929100}

\newcommand{\substepseparator}{\hspace{1cm}}

\title{Steps Before Syntax: Helping Novice Programmers Solve Problems using the PCDIT Framework}

\author{Oka Kurniawan\textsuperscript{1}, Cyrille Jégourel\textsuperscript{1}, Norman Tiong Seng Lee\textsuperscript{1}, Matthieu De Mari\textsuperscript{1}, and Christopher M. Poskitt\textsuperscript{2} \\
  \textsuperscript{1}Singapore University of Technology and Design, Singapore \\
  \textsuperscript{2}Singapore Management University, Singapore \\
  {\underline{ \{oka\_kurniawan, norman\_lee, cyrille\_jegourel, matthieu\_demari\}@sutd.edu.sg}, \underline{cposkitt@smu.edu.sg}} \\}

\date{}

\begin{document}
\maketitle
\begin{abstract}

Novice programmers often struggle with problem solving due to the high cognitive loads they face. Furthermore, many introductory programming courses do not explicitly teach it, assuming that problem solving skills are acquired along the way. In this paper, we present `PCDIT', a non-linear problem solving framework that provides scaffolding to guide novice programmers through the process of transforming a problem specification into an implemented and tested solution for an imperative programming language. A key distinction of PCDIT is its focus on developing concrete cases for the problem early without actually writing test code: students are instead encouraged to think about the abstract steps from inputs to outputs before mapping anything down to syntax. We reflect on our experience of teaching an introductory programming course using PCDIT, and report the results of a survey that suggests it helped students to break down challenging problems, organise their thoughts, and reach working solutions.
\end{abstract}


\section{Introduction}

Learning programming from scratch is understood to be challenging~\cite{Kelleher2005}. This is mainly due to the high cognitive load involved in typical introductory courses: novice programmers must learn language syntax, IDEs, engineering principles (e.g.~abstraction, modularity), and computational thinking all for the first time. Until their mental models have fully developed, novices can struggle to get started on a harder programming problem, sometimes taking a `syntax-first' approach of coding something---\emph{anything}---before even analysing how to properly solve it. This is compounded by the fact that many introductory courses do not explicitly teach \emph{problem solving} skills and strategies, assuming instead that they are picked up along the way~\cite{DeRaadt2004}.

Several authors have developed approaches for guiding novice programmers through the process of problem solving. For example, in their Python textbooks, Dierbach~\cite{Dierbach2013} and Liang~\cite{liang2012a} proposed frameworks based on the steps of the software development life cycle, i.e.~analysis/requirements, design, implementation, and testing. Students are encouraged to think about the input/output requirements of the problem, ``design a process for obtaining the output from the input''~\cite{liang2012a}, implement it, then finally ``test the program on a selected set of problem instances''~\cite{Dierbach2013}. As another example, Loksa et al.~\cite{Loksa2016} proposed a framework that expands upon the design phase by encouraging students to search for analogous problems and solutions that can be applied to the one they are tackling.

These approaches share a typical characteristic in that testing is at the end of the process. Actually being able to get to the end, however, depends on the student's level of metacognitive awareness, i.e.~their ability to think on their own about the problem~\cite{Polya1945}. For students lacking metacognitive skills, previous research has highlighted the benefit of solving concrete cases \emph{before} programming~\cite{Denny2019}, as well as making the problem solving process explicit (e.g.~by using an automated assessment tool) and having them reflect on their progress~\cite{Prather-et_al18a,Prather-et_al19a}. Controlled experiments in these works revealed that students are more likely to provide a correctly implemented solution to a problem and demonstrate better metacognitive awareness.

\begin{figure}[!t]
  \centering
  \includegraphics[width=0.7\linewidth]{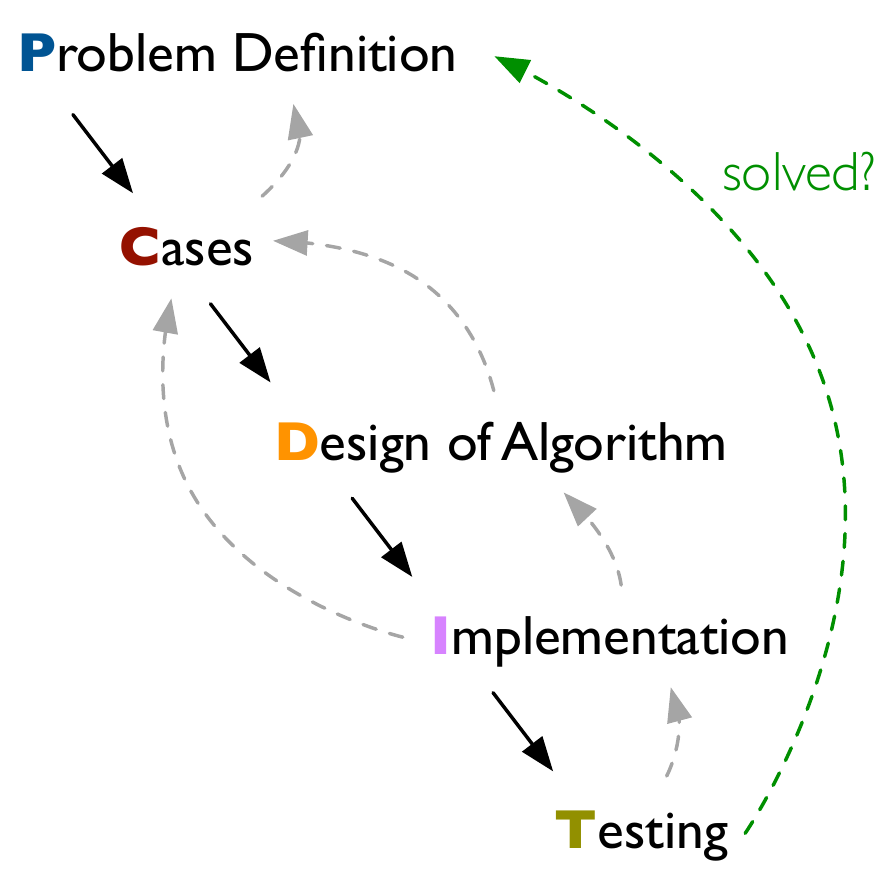}
  \caption{Steps of the PCDIT framework, including some possible non-linear flows between them}
  \label{fig:pcdit_figure}
\end{figure}

It is in this context that we developed `PCDIT', a problem solving framework for novice programmers that encourages them to design/solve concrete cases before programming, and to regularly reflect using the five eponymous phases of the process (Figure~\ref{fig:pcdit_figure}): \textbf{\textcolor{MyP}{P}}roblem Definition, \textbf{\textcolor{MyC}{C}}ases, \textbf{\textcolor{MyD}{D}}esign of Algorithm, \textbf{\textcolor{MyI}{I}}mplementation, and \textbf{\textcolor{MyT}{T}}esting. A key characteristic of the framework is the `C' phase, which focuses on developing concrete cases for the problem early without actually writing any test code: students instead think about the steps at an abstract level, only mapping them down to program syntax in later phases. Another characteristic is its non-linearity: students are encouraged to engage in a reflective, case-driven, and iterative process that may feel more productive and encouraging for novices~\cite{Rist1989}. The process keeps on going until the programming task has been solved satisfactorily, meaning that the proposed answer fulfills all the requirements described in the problem statement, and can operate correctly on all test cases. Ultimately, the scaffolding of PCDIT makes the process of problem solving that we naturally follow---but many novices do not---explicit.

In this experience report, we introduce the PCDIT framework, and describe how we used it in a first-year undergraduate programming course, where it was explicitly taught as a way to help solve more challenging coding exercises. We present the results of a post-use student survey that suggests it was successful in helping students to organise their thoughts, break down problems, and arrive at working solutions. We critically reflect on our experiences---as software engineering lecturers---of using it in teaching, and make a number of recommendations on how to do so more effectively. Finally, we share a PCDIT worksheet and a number of examples that can be used by other practitioners.

\section{Related Work}
\label{sec:related_work}

In the mathematical problem solving framework described in Pólya's book, ``How to Solve It''~\cite{Polya1945}, the reader is told to: (1)~understand the problem, (2)~devise a plan, (3)~carry out the plan, and (4)~look back (i.e.~evaluate the solution).  In the second stage, the reader is asked ``Have you seen it before? Do you know a related problem?'', implicitly suggesting that successful problem solving requires an adequate knowledge base or cognitive resources. On top of this, according to Schoenfeld~\cite{Schoenfeld1985}, successful problem solving requires three other aspects in the learner: (1)~heuristics (strategies for making progress in unfamiliar situations), (2)~control (making decisions about strategies and resources), and (3)~beliefs (i.e.~about the subject). 

Some introductory Python programming textbooks~\cite{Dierbach2013,liang2012a} introduce problem solving using a framework that is based on the software development life cycle and is analogous to Pólya's, namely: (1)~requirements and analysis, (2)~design, (3)~implementation, and (4)~testing. In the first stage, Liang~\cite{liang2012a} tells the reader to use the requirements identified to determine the input and output data for the problem. Dierbach~\cite{Dierbach2013} points out that the reader also needs to consider how the input data can be represented in the program itself. For the remaining stages, both authors then tell the reader to write the algorithm, implement it in code, and then evaluate the solution using test cases. While both books use worked examples to illustrate concepts, Dierbach's book also has specific sections that illustrate the problem solving process throughout the chapters. 

Other authors adopt a framework that is similar to Pólya's, but with emphasis on different aspects. McCracken et al.~\cite{McCracken2001} highlights a stepwise refinement strategy, citing the need to decompose the problem into sub-problems with the following framework: (1)~abstract the problem from its description, (2)~generate sub-problems, (3)~transform sub-problems into sub-solutions, (4)~recompose, and finally, (5)~evaluate and iterate. Loksa et al.'s framework~\cite{Loksa2016} tells the reader to draw upon a knowledge base of existing problems in stages (2)--(4): (1)~reinterpret problem prompt, (2)~search for analogous problems, (3)~search for solutions, (4)~evaluation of a potential solution, (5)~implement a solution, and (6)~evaluate implemented solution. 

Some authors propose that, at the problem analysis stage, examples should be written to illustrate the purpose of the program. Riley~\cite{Riley1981} describes requiring students to write a problem definition which contains a problem description, input and output specifications, error conditions, and specific examples that illustrate the input and outputs.  In the six-stage process proposed in ``How to Design Programs''~\cite{felleisen2018}, the reader is told to provide some functional examples in the initial stages of problem analysis, prior to carrying out any design. This involves providing specific pairs of input and output values in order to illustrate the purpose of the function to be written.

Studies have been done on the problem solving ability of students in first-year programming courses. While McCracken et al.~\cite{McCracken2001} suggested that novice learners of programming have no problem solving ability, subsequent studies have concluded that reducing students' cognitive load might have a positive effect on problem solving performance. In a study by McCartney et al.~\cite{McCartney2013}, participants were provided partially-completed code and allowed access to external materials, thus reducing their cognitive load. Many participants demonstrated the ability to program incrementally. Utting et al.~\cite{Utting2013} showed that participants who were provided with a test harness had better success in completing an object-oriented programming task, compared to those who had to rely solely on the program description.

It has been pointed out that novice learners of computing have ``fragile knowledge''~\cite{Perkins1986}. This was illustrated in a study by Lister et al.~\cite{Lister2004}, where novice learners were shown to have a weak ability to read and trace computer programs, lacking the prerequisite skills for problem solving. Thus, pedagogical strategies have been proposed to help novice programmers develop their problem solving skills. 

One pedagogical strategy has been to teach students explicitly algorithmic solutions to known problems, e.g.~a counter-controlled loop, termed as ``goals and plans'' by Soloway~\cite{Soloway1986} and ``algorithmic patterns'' by other authors~\cite{Muller2005, Ginat2004}. Soloway points out that these must be explicitly taught to students. In the problem solving process, students must apply a stepwise refinement process and break down a problem into sub-problems in such a way that these known patterns can be used~\cite{Soloway1986}. Clancy and Linn~\cite{Clancy1999} point out that regular classroom instruction must continually focus on helping students to learn and apply these patterns.

A curriculum that used Soloway's goal/plan framework and  explicitly taught 18 algorithmic patterns was reported by de Raadt~\cite{deRaadt2009}. The study showed that students successfully used these strategies more often, compared to previous iterations of the course that only exposed students to these patterns implicitly.
 
Specific instruction in class on problem solving has also been used as a pedagogical strategy. Arshad~\cite{Arshad2009} used recitation sessions where teaching assistants demonstrated solutions to programming problems and articulated their thoughts aloud at the same time. Students reported that this aspect of the course was most useful for them. 

Falkner and Palmer~\cite{Falkner2009} described a class with three interventions: live demonstrations of programming examples, lessons that specifically discussed problem solving, and encouraging cooperative problem solving in students. 

Loksa et al.’s study~\cite{Loksa2016} seemed to show that interventions designed to improve problem solving skills and encourage metacognition had a positive effect on novice learners. Participants in the study were explicitly taught the problem solving framework. As the participants worked on the programming tasks and encountered difficulties, they were prompted to think about which stage of the problem solving framework they were in. The IDE used by the participants had an ``idea garden'', which was a set of problem solving strategy hints which students could use. 

Prather et al.~\cite{Prather-et_al18a} described metacognitive difficulties that students faced, noting that many students misunderstood the problem prompt and formed the wrong conceptual model, began coding straightway without designing the solution, and were not able to respond correctly to feedback from error messages and failed test cases. They also pointed out that current automated assessment tools~(AAT) did not have features that facilitated students through Loksa et al.'s problem solving framework. 

Subsequently, an AAT was modified to require students to solve a randomly generated test case after being presented with the problem  \cite{Denny2019,Prather-et_al19a}. Only upon solving this test case are they allowed to begin implementing their solution. This was based on the hypothesis that this would facilitate students' metacognitive awareness in the earlier problem solving stages. It was shown that students who worked on the test cases before their implementation showed better problem solving outcomes compared to a control group. 

There have also been AATs designed to require students to input test cases together with their solution, forcing them to demonstrate the correctness of their solution. In the AAT used by Edwards~\cite{Edwards2004}, students are graded both on their test suite and their solution. Wrenn and Krishnamurthi~\cite{Wrenn2019,Wrenn2020} augmented their AAT with a tool for students to write test cases and evaluate their thoroughness and completeness, which is done prior to implementing the solution to the problem. There was evidence that students used this tool prior to implementation even when not required, suggesting a change in student behaviour as a result of this tool.

\section{Context}
\label{sec:context}

The motivation to develop a problem solving framework for novice programmers came from teaching \emph{``Digital World''}, a first-year undergraduate programming course at the Singapore University of Technology and Design. Our institution primarily offers four-year engineering degree programmes, which students apply to after completing post-secondary education. A unique characteristic of our programmes is that they share a common first year, meaning that \emph{every} student is required to take our course, regardless of what they ultimately major in.

Our course covers the fundamentals of computational thinking and programming using Python, and is designed to be accessible to students without any prior experience in the language, taking them from the basics (variables, types, conditionals) through to some introductory object-orientation material. It is delivered using elements of the flipped classroom~\cite{Lage-Platt-Treglia00a,Mok14a,Maher-et_al15a}: before coming to class, students are expected to read some instructor-prepared materials and complete a reading quiz, so that they can be primed and ready to focus on exercises and deeper technical discussions during class. Our materials are hosted on Google Colab~\cite{Nelson-Hoover20a} (a hosted Jupyter Notebook~\cite{Willis-et_al20a} interface), which allows for code snippets to be run in the browser and thus makes examples more interactive.

As the course is taken by every undergraduate student at our institution, it caters to a wide range of abilities and backgrounds in programming---including none. In previous years, we found that novice programmers were often struggling with where to start on any exercise that went beyond the very basics. Novice programmers would often take a `syntax-first' approach, where they would type \emph{some} code that they saw earlier without really thinking about what it does, or whether it takes any steps forward towards the solution, eventually leading to tangled-up code that has them demotivated and hitting a brick wall. Observing this pattern year after year led us to design the problem solving framework of this paper: we wanted to encourage such students to reflect, organise their thoughts, think through the problem at an abstract level, and only look for the right syntax/constructs once the problem and solution are clear.

\section{Our Intervention: PCDIT}
\label{sec:interventions}

\noindent\textbf{PCDIT Framework.} Our pedagogical intervention was the design of the PCDIT framework and the decision to explicitly teach it in our introductory programming course. The five key steps of the framework are given in Figure~\ref{fig:pcdit_figure}, and are intended to capture the problem solving process that we (as instructors) naturally follow, but that novice programmers are not yet familiar with. By naming the steps and providing an acronym, we make it easier for students to reflect on where they are in the problem solving process, potentially increasing their metacognitive awareness~\cite{Prather-et_al18a,Prather-et_al19a}. It is important to note that instructors and students are encouraged to go iterate between the steps as their thinking brings clarity.

In general, the process begins with forming a \textbf{\textcolor{MyP}{P}}roblem Definition: students are asked to identify the types of inputs and outputs and summarise in natural language what it is that needs to be solved (e.g.~``take a single string value as input, then return the reverse of that string as output''). This step is common to many problem solving frameworks in understanding the problem and formulating it. Similar to Dierbach~\cite{Dierbach2013}, students are encouraged to provide more detail on the kind of data involved in both the input and the output and how it can be represented in the program. This step also requires students to summarise the problem in a single statement, similar to Riley's `general description'~\cite{Riley1981}.

The second phase asks students to develop concrete \textbf{\textcolor{MyC}{C}}ases, i.e.~before even thinking about the algorithm. The intention is for students to conceptualise the abstract steps from concrete inputs to outputs, helping them to generalise to an algorithm more easily in later parts of the framework. The \textbf{\textcolor{MyC}{C}}ases step is similar to the functional examples in ``How to Design Programs''~\cite{felleisen2018}. As students work on various concrete cases, they can also step back and revise their problem definitions, e.g.~adding additional information about the required data types. This step is crucial for novice programmers, many of whom do not have any existing algorithmic patterns or schemas: it may be difficult for such students to search for analogous problems as in Loksa's framework~\cite{Loksa2016}. They need to build the solution from the bottom up and this concrete \textbf{\textcolor{MyC}{C}}ases step provides a bridge to figure out the algorithmic solution in the next step. As discussed in Section~\ref{sec:related_work}, working out specific concrete cases helps students to understand the problem better, and there is evidence it helps them in implementing their solutions. We highlight that while some frameworks encourage students to write cases as part of their testing code, in PCDIT, they focus on \emph{working out the abstract steps} (e.g.~on paper) from concrete inputs to outputs.

\begin{figure}[!t]
  \includegraphics[width=\linewidth]{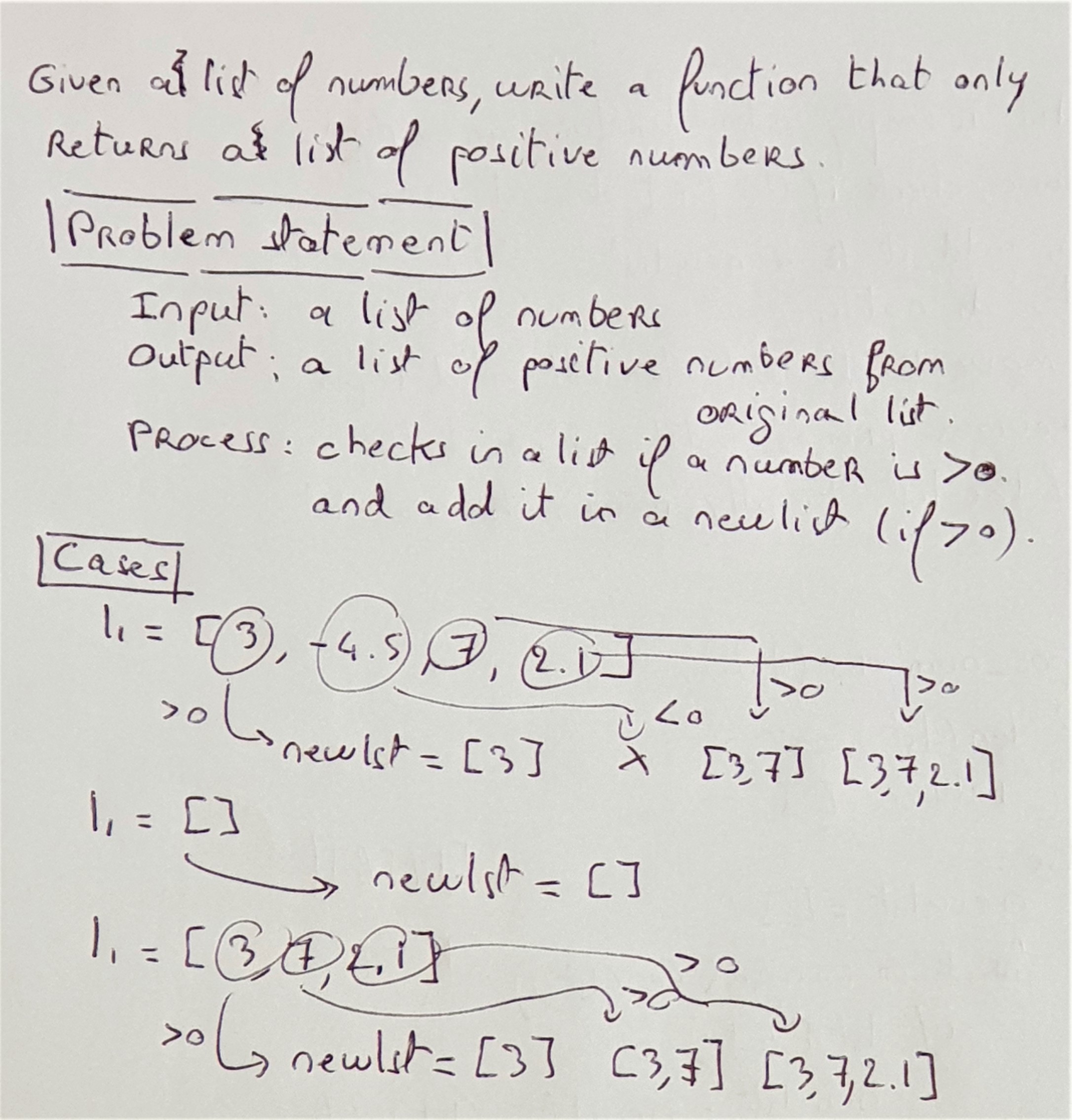}
  \caption{`P' and `C' steps examples for list of positive numbers}
  \label{fig:annotated_worksheet}
\end{figure}

\begin{figure}[!t]
  \includegraphics[width=\linewidth]{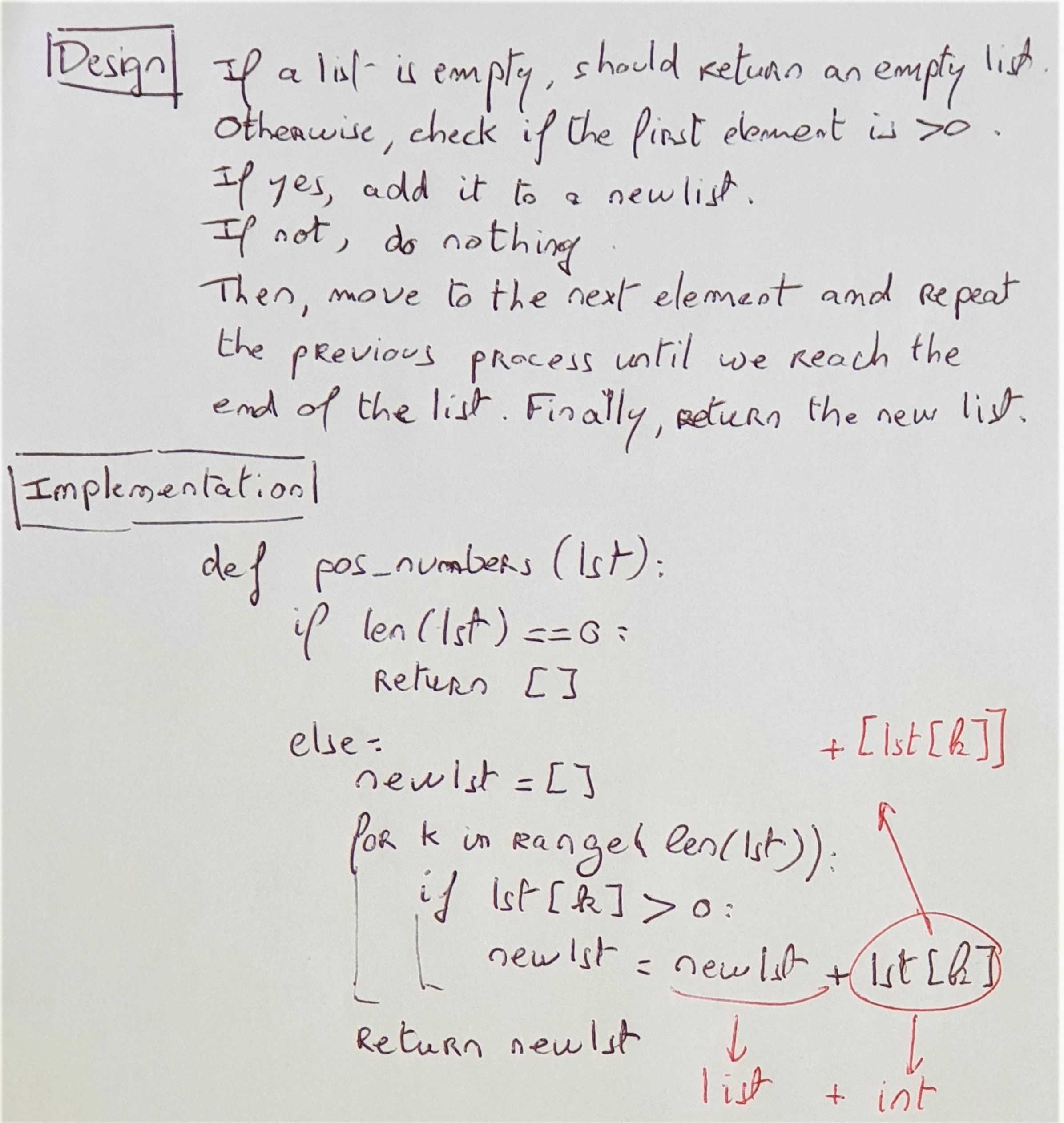}
  \caption{`D' and `I' steps examples for list of positive numbers. The initial `I' step in this example contains an error, discovered in the `T` step}
  \label{fig:annotated_worksheet_DI}
\end{figure}

Once students have worked on these concrete cases, they can begin the \textbf{\textcolor{MyD}{D}}esign of Algorithm phase: for each concrete input/output, we ask them to enumerate the steps they did in working out the concrete \textbf{\textcolor{MyC}{C}}ases. They are asked to look back on how they arrive at the output starting from the input. We then ask them to identify patterns in those steps and generalise them to computational steps. These steps can be written in a mix of pseudo-code and (precise) natural language---whatever the student is currently more comfortable with. This part can be iterated several times, starting with more coarse subgoals/descriptions, before refining them over the iterations, e.g.~by employing specific \emph{key words/phrases}, such as ``for every element in\dots'', ``as long as\dots'', or ``compare if\dots''. Using specific keywords that sound similar to programming language syntax eases the transition from pseudo-code to actual code later. Figures~\ref{fig:annotated_worksheet}--\ref{fig:annotated_worksheet_DI} illustrate how one can take some concrete \textbf{\textcolor{MyC}{C}}ases for a problem and start to sketch an algorithm \textbf{\textcolor{MyD}{D}}esign in an intuitive (but not yet fully refined) way. (Further examples are available in Examples 1--3 of~\cite{pcdit_additional}.)

In the subsequent steps, students start to map the pseudo-code of their solution down to concrete Python in the \textbf{\textcolor{MyI}{I}}mplementation phase. In our teaching, we iterate this part of the framework with the \textbf{\textcolor{MyT}{T}}esting phase, ensuring that students are regularly testing their programs after completing every few lines of mapping. This helps to ensure novices feel motivated and productive by tackling smaller/feasible sub-problems one-by-one. In testing their code, students can use some of the \textbf{\textcolor{MyC}{C}}ases identified earlier, or propose new ones that potentially highlight the need to go back and improve other aspects of the algorithm further. The \textbf{\textcolor{MyI}{I}}mplementation step in Figure~\ref{fig:annotated_worksheet_DI} actually contains a syntax error in the list operation, illustrating the importance of going back and revising the initial implementation after the \textbf{\textcolor{MyT}{T}}esting step.

We created a PCDIT worksheet that we share with students when teaching the framework~\cite{pcdit_additional}. Among our supplementary materials, we include fully worked examples (matrix multiplication, extraction of positive numbers of a list, etc.) using our PCDIT worksheet~\cite{pcdit_additional}.

\substepseparator

\noindent\textbf{Implementation.} We now describe how the framework was taught in our first-year programming course (Section~\ref{sec:context}), and how we elicited the reflections of the students using an optional survey.

The instructors explicitly introduced the PCDIT framework in the third or fourth week of the course, when students began to see more complicated problems, e.g.~dealing with loops and a combination of conditionals and iterations, together with the string, list, and dictionary data types. The problems that the instructors used to demonstrate the PCDIT framework involved the synthesis of several concepts such as: (1) manipulating strings using loops and conditionals, and (2) operations involving dictionaries or lists. Fully worked examples of these problems can be found in our supplementary material~\cite{pcdit_additional}.

In the lesson segment involving the PCDIT framework, after a brief introduction, instructors would show the problem to the students, and show how the PCDIT framework should be used by filling up the worksheet and displaying the process on the projector. Students were reminded that writing the code would only begin \emph{after} some iterations through the first three stages had been satisfactorily completed. The segment then ended with a live coding demonstration showing how the completed PCDIT worksheet could be used in conjunction with the code that was written. 

At the end of the lesson segment, an instructor from a different class came to invite students to fill in a survey on their experience of applying PCDIT. The survey was done online and a follow-up interview was available for those who were willing to take part. The survey asked them about their confidence in problem solving and their computing self-efficacy~\cite{Gok2011DevelopmentOP, Kolar2013}. We also asked whether the PCDIT framework helped them in their problem solving process and in writing Python code.

\section{Survey Results}
\label{sec:survey_results}

\begin{figure*}[!t]
  \centering
  \subfigure[Questions on interest in problem solving]{\includegraphics[width=0.4\textwidth]{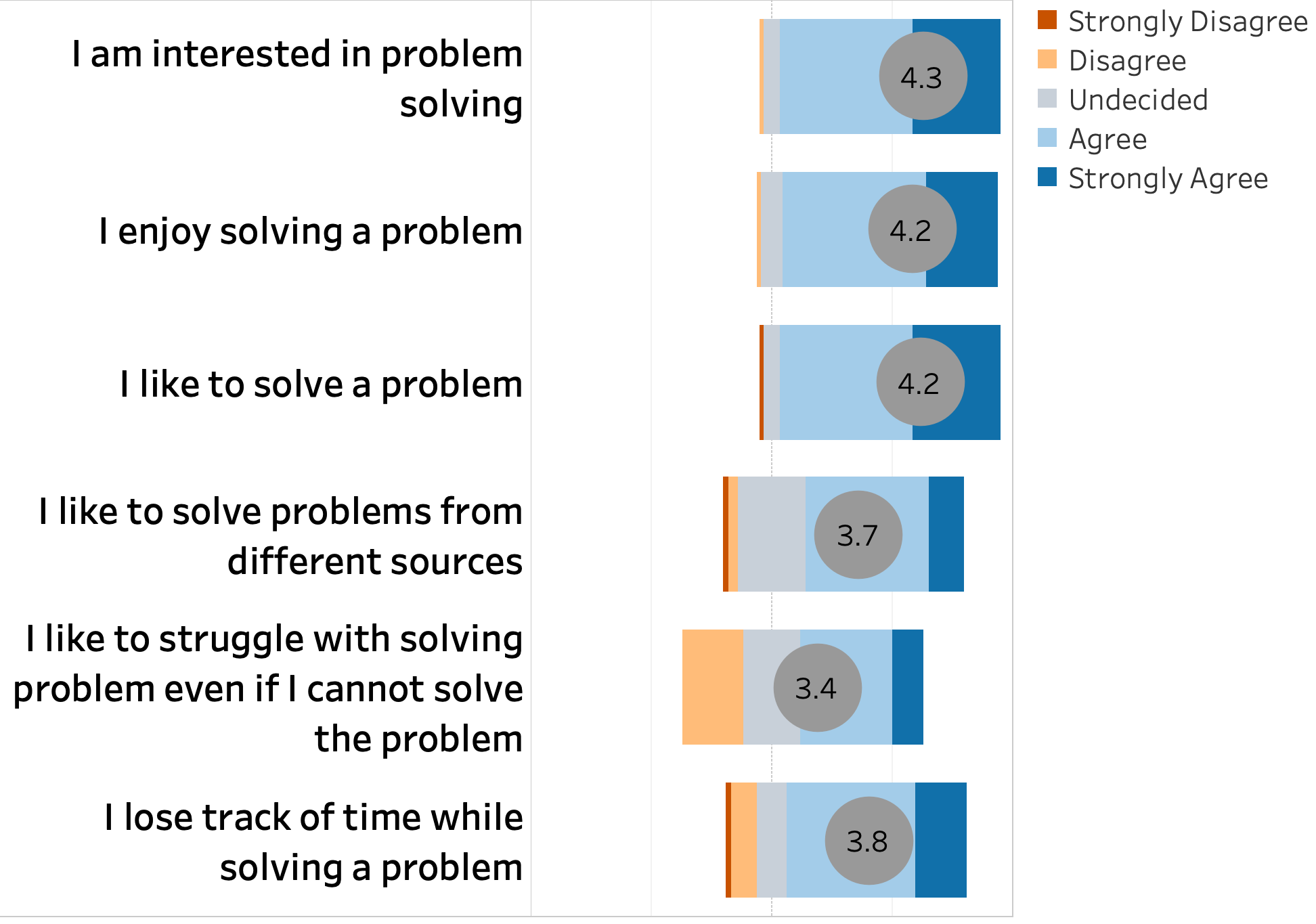}}\quad
  \subfigure[Questions on effort to solve problems]{\includegraphics[width=0.4\textwidth]{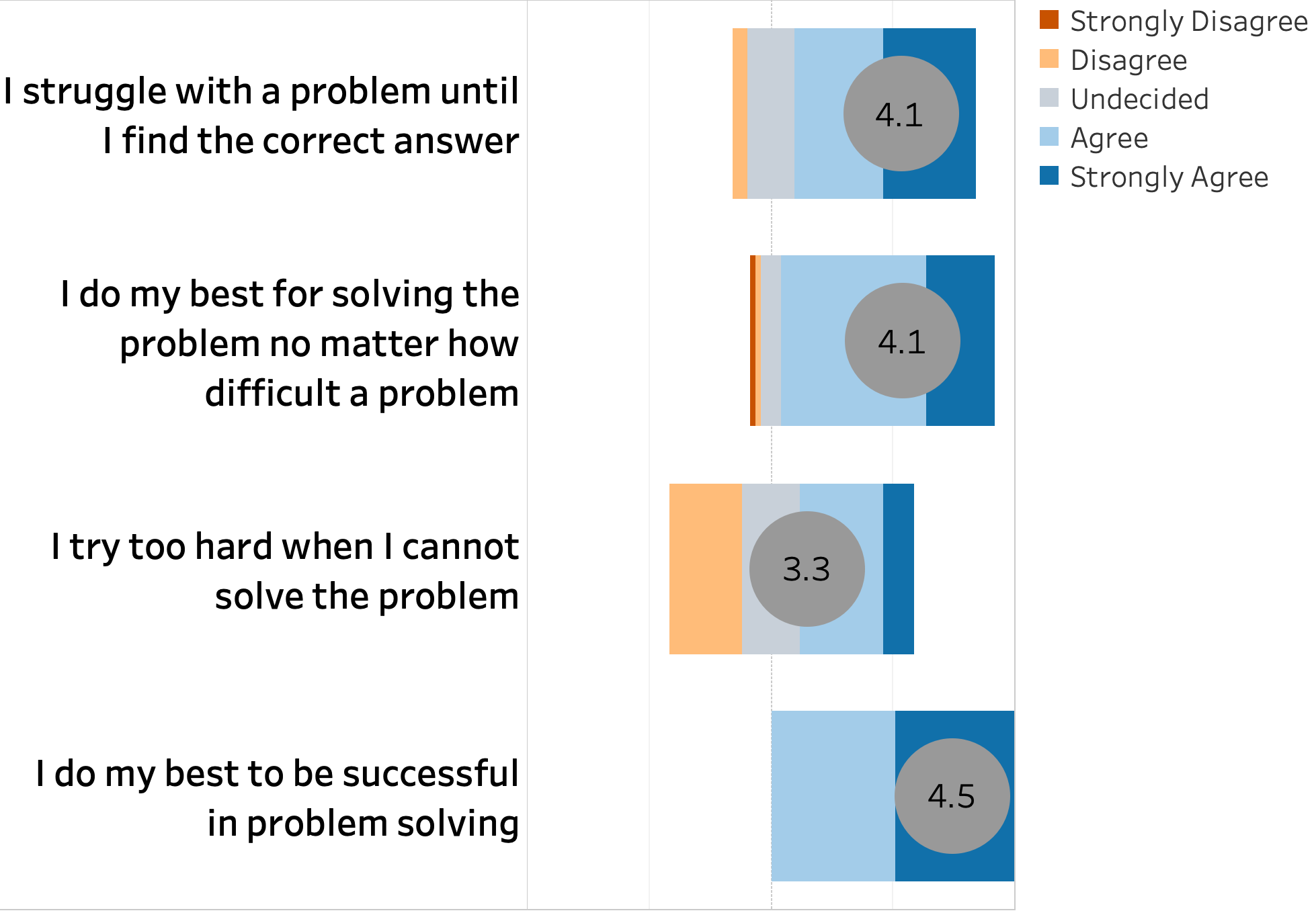}}\quad
  \subfigure[Questions on confidence to solve problems]{\includegraphics[width=0.4\textwidth]{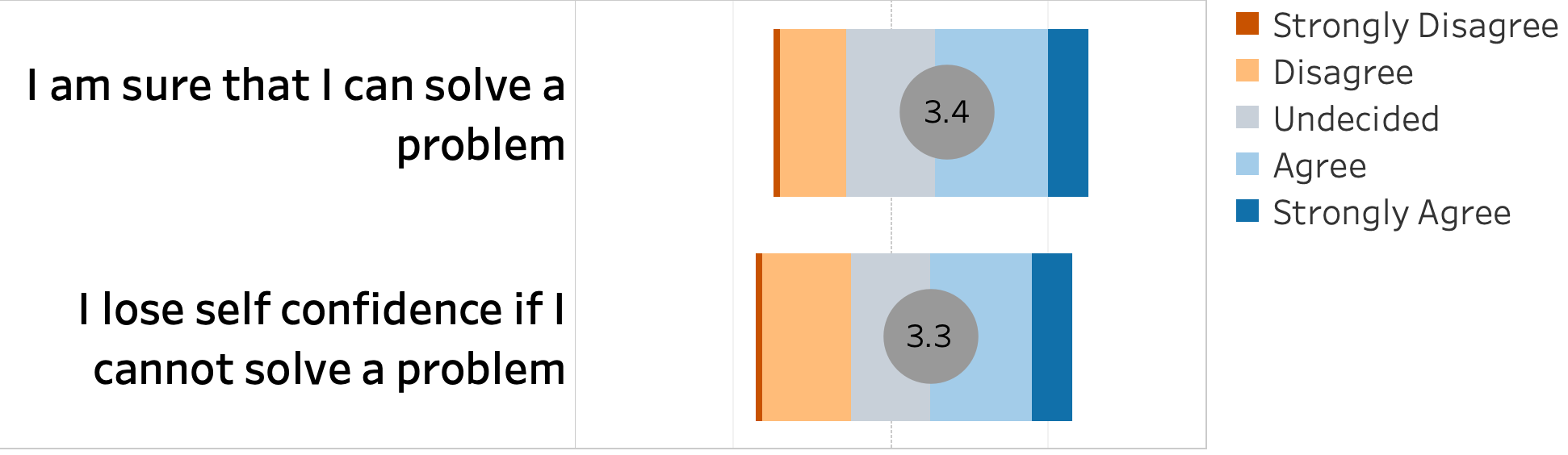}}\quad
  \subfigure[Questions on the helpfulness of the PCDIT framework]{\includegraphics[width=0.4\textwidth]{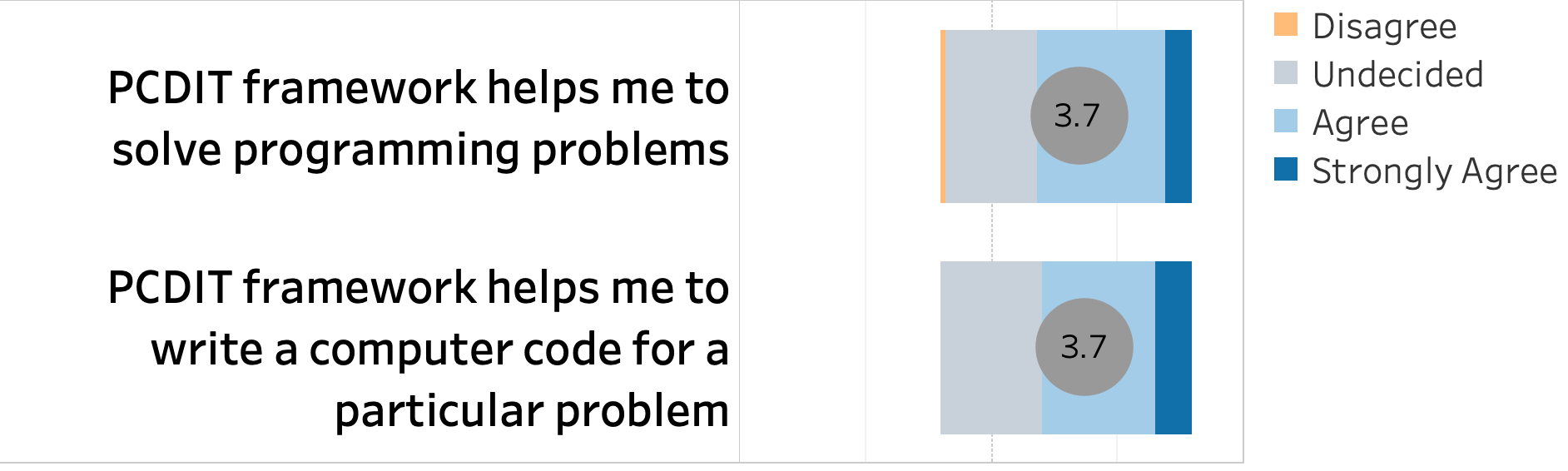}}
  \caption{Results (5-point Likert scales) from our post-use student survey on problem solving and PCDIT}
  
  \label{fig:survey_results}
\end{figure*}

Our optional survey elicited 47 responses from three classes, indicating a response rate of about $35\%$. Of these, 25 students were female and 22 were male. Before joining our university, the majority of respondents ($66\%$) had studied at junior colleges, i.e.~institutions that offer pre-university courses such as the GCE Advanced Levels or the International Baccalaureate Diplomas. Other students had studied at polytechnics ($12\%$) and international schools ($20\%$). A number of students ($38\%$) had previously written more than 500 lines of code, whereas $40\%$ had written between 50-500 lines, $18\%$ had written between 1-50, and $3\%$ had never written any. Only eight students explicitly reported using Python (the language of our course) in the past; other students reported some experience in C++, Java, and web programming languages (e.g.~HTML, JavaScript). Finally, $49\%$ of the students indicated that they were interested in majoring in our information systems or computer science programme, i.e.~likely indicating an explicit interest in programming.

The results in Figure \ref{fig:survey_results} indicate that the survey participants show positive interest towards problem solving. The scores in Figure \ref{fig:survey_results}(a) tend to be above 3.0 (on a 5-point Likert scale). The results also correlate on the effort that they would put in when solving a problem. Figure \ref{fig:survey_results}(b) strongly suggests that most participants put in effort when faced with problem solving tasks. On the other hand, the survey result on their confidence level in solving a problem is not as high as their interest and effort. It can be seen from Figure \ref{fig:survey_results}(c)  that the overall Likert average score is only slightly above 3 (3.4) for the positive statement ``I am sure I can solve a problem''. The negative statement ``I lose self confidence if I cannot solve a problem'' also results in a score of 3.3 which shows the ambivalence of the respondents' self-confidence in solving problems. 

Figure \ref{fig:survey_results}(d) suggests agreement that the PCDIT framework, after being explicitly taught in class, helped the respondents to solve programming problems and write code. Both of the average Likert scores were about 3.7 out of 5.0. We did not see any respondents choosing score 1 (Strongly Disagree) and saw only one student choosing score 2 (Disagree). This shows that majority of respondents (62\%) find the framework helpful in their problem solving and in mapping the solution to code.

\begin{figure}[!t]
\centering
\includegraphics[width=0.9\linewidth]{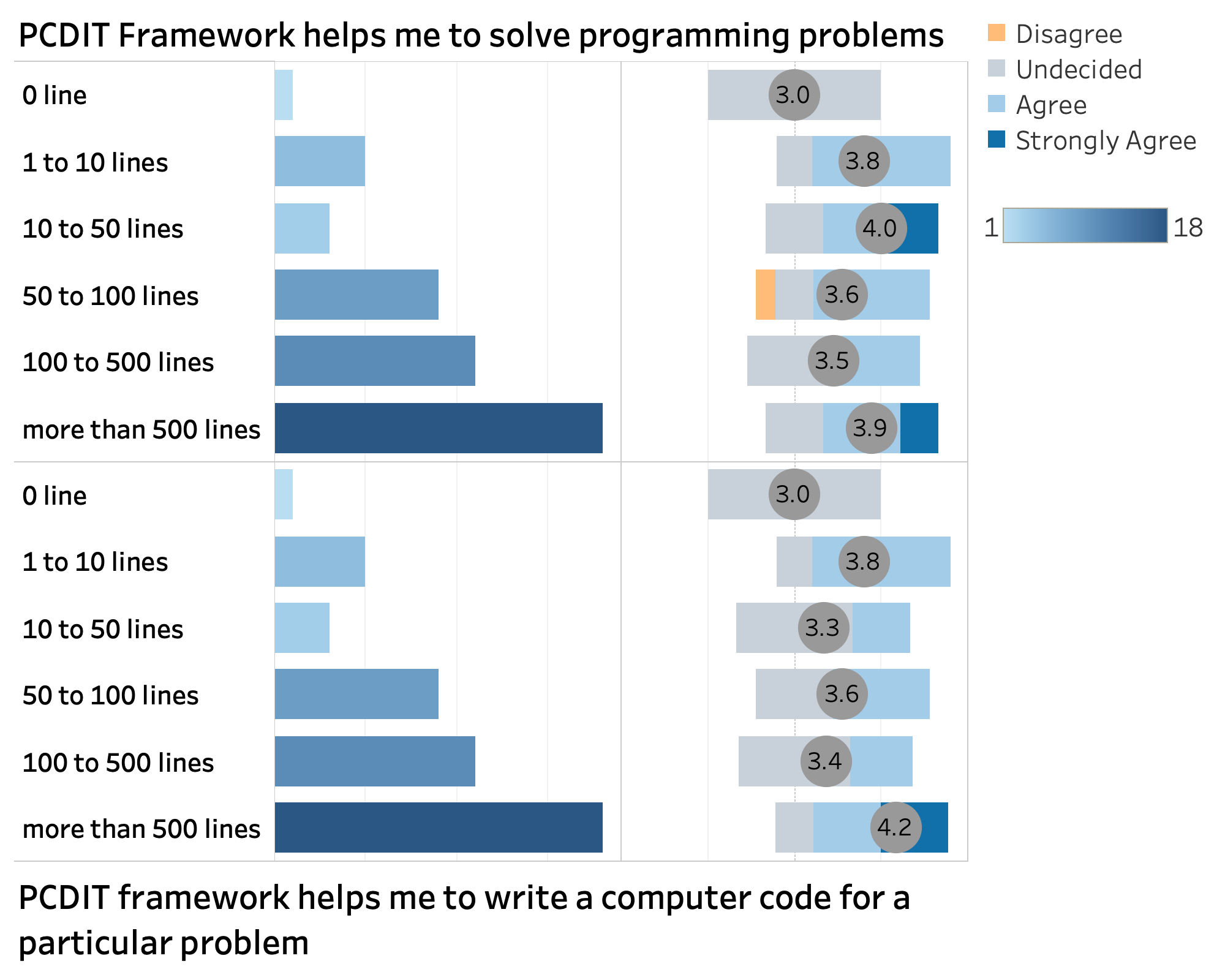}  \caption{Survey results (5-point Likert) regarding PCDIT plotted against pre-course coding experience}

  \label{fig:pcdit_background}
\end{figure}

In Figure~\ref{fig:pcdit_background}, we break down the results further by plotting the PCDIT responses against the respondents' programming backgrounds (i.e.~lines of code written before the course). There was only one student who had written 0 lines in the background survey, and that student chose ``Undecided'' for whether the framework is helpful. Other respondents leaned towards agreement or strong agreement. We also observed an interesting result in that there were those who had written more than 500 lines and yet still continued to strongly agree that the framework helps them in solving programming problems and writing the code.

We also asked students what they found most useful from the framework. A number of students commented on its systematic nature and step-by-step approach. They also mentioned that the framework helps their thinking or thought process. In fact, the pause it forces before writing the code actually helps them in solving the problem itself. Below are a few quotes from the students' comments:

\begin{quote}\footnotesize
    \emph{When you write out the thinking process it helps to solve the problem.}
\end{quote}
    
\begin{quote}\footnotesize
    \emph{The pause it forces me to take before solving something; sometimes I can be too anxious and jump to conclusions without having thought carefully about the problem.}
\end{quote}

Another point that we see from the comments is related to \textbf{\textcolor{MyT}{T}}esting. Quite a number of comments indicated that the framework helps them to better test the code, as shown by the following quotes.

\begin{quote}\footnotesize
    \emph{Check potential problems.}
\end{quote}

\begin{quote}\footnotesize
    \emph{Testing codes using different test cases.}
\end{quote}
    
We also asked what they found to be least helpful with regards to this framework. Most of the respondents indicated that it is time consuming. One response explicitly said that such a step may be good for beginners but not for experienced programmers.

\begin{quote}\footnotesize
    \emph{It's good for beginners, but may be time consuming for experienced programmers.}
\end{quote}

One response commented that they do not know the difference between the \textbf{\textcolor{MyC}{C}}ases step and the \textbf{\textcolor{MyT}{T}}esting step as seen from the quote below.

\begin{quote}\footnotesize
    \emph{C:Test cases. Is this not also a part of T:Testing?}
\end{quote}

We will take up the survey results, students' comments as well as instructors' reflection in the following section for our reflection and discussion.

\section{Critical Reflections}
\label{sec:reflections}

In this section, we reflect and discuss our experiences of using the PCDIT framework to teach students problem solving in our programming course.  First, we reflect on the viewpoints of students who used the framework. Next, we discuss the reflections of instructors who facilitated it during class.

\substepseparator

\noindent\textbf{Student Reflections.} The survey respondents had varied programming backgrounds. Quite a number of them had actually written some code in the past, with only 20\% having written 50 lines or fewer. This likely explains the strong positive scores for interest and effort in our questions. Despite this, we noticed some ambivalence with respect to confidence in solving problems. This means that even though the students tend to enjoy solving problem and have some experience in writing computer code (more than 30\% had written more than 500 lines of code), their confidence level in solving a problem is not high. This shows awareness from the students that being able to solve problems is not the same thing as simply writing code.

The breakdown shown in Figure~\ref{fig:pcdit_background} gives some insight into how students' programming background may affect their perception of the PCDIT framework. As mentioned, no student chose `Strongly Disagree' and only one student chose `Disagree' when asked whether the framework helps them. In general, the majority of students agreed or strongly agreed that the framework helped them. More students chose `Undecided' as a response for the question on whether the framework helps them to write computer code as compared to whether the framework helps them in solving programming problems. This suggests that the framework helps more on thinking through the problem solving process. This also could mean, that for most students, they need more help on the problem solving process rather than on finding the right syntax to use. 

What is interesting, however, is that the majority of students who had written more than 500 lines of code found the framework helpful. These are students with some programming experience and who may have learnt Python syntax previously. This again agrees with the observation that the framework's main contribution is on the problem solving process. For these students, mapping a solution to Python code seems not to be one of their main challenges. This also could be the reason why more of them voted favourably about the framework.

These conclusions are strengthened further when taking the students' written comments into account. Most students indicated that the framework helps them in organising their thoughts, in the process of thinking through the problem, and in making their thinking process more systematic. None of them mentioned that it helps them to actually write the Python code. So it is more on the process of problem solving that the framework appears to contribute.

On the other hand, the very same advantage provided by the framework can also be considered one of its disadvantages: the thinking process and the stages of the framework \emph{consume more time}. In the context of a synchronous class or exam, students may be reluctant to use such a method. One way to speed up the application of PCDIT would be to write both the \textbf{\textcolor{MyP}{P}}roblem Definition and \textbf{\textcolor{MyD}{D}}esign of Algorithm steps in the editor or AAT itself, e.g.~as part of the code's comments.

One of the students who volunteered for a follow-up interview gave further insight on how students with experience in programming use this framework. The student recognised that these steps are some of the things he subconsciously does when solving a problem. The framework helps to make the thinking process more explicit and clear. Moreover, he also stated that he found it useful for more complicated problems where the solution is not a one-liner program. This agrees with the purpose of the framework which is to help with the problem solving difficulties that students face when solving more complex exercises. Another interesting point from the interview was that he actually went through the PCDIT process when doing his \emph{group} programming assignment. They started by formulating the problems. Then, they tried to identify the functions by trying to work out the different cases. He explicitly said that the framework was used to facilitate a kind of group discussion or brainstorming process for solutions. This shows that the framework can be used not only by individuals but also as a group in a collaborative work. Further study on how this framework can be applied in a group setting should be explored.

\substepseparator

\noindent\textbf{Instructor Reflections.} Three instructors involved in teaching the PCDIT framework were asked to share their reflections. All three found that the framework is natural and makes explicit what it is they naturally do when teaching programming. For example, one of them said:

\begin{quote}\footnotesize
    \emph{I realized that many parts of PCDIT  are implicitly built into the problem sets that I designed.}
\end{quote}

Another mentioned that the PCDIT structures what it is he does when solving problems himself:

\begin{quote}\footnotesize
    \emph{The framework is well structured and the acronym seems easy enough to be remembered by the students. I have also realised that the PCDIT approach is actually more or less the approach I am using in practice.}
\end{quote}

With respect to the difficulties in teaching using the framework, most of them identified the second step---working on the \textbf{\textcolor{MyC}{C}}ases---to be one that students find challenging, and that more time should be spent during class to cover it. One of them stated that most students do not see the difference between the \textbf{\textcolor{MyC}{C}}ases and \textbf{\textcolor{MyT}{T}}esting stages and more emphasis on their difference should be done in the teaching. One example on how \textbf{\textcolor{MyC}{C}}ases and \textbf{\textcolor{MyT}{T}}esting stages differ can be found in Example 3 in \cite{pcdit_additional}. In this example, the \textbf{\textcolor{MyT}{T}}esting step helps to identify an error when adding an item into a list which is particularly dependent on the programming language syntax and features. On the other hand, the \textbf{\textcolor{MyC}{C}}ases step guides the design of algorithm and would not be able to detect such mistakes. Example 4 in \cite{pcdit_additional} shows how \textbf{\textcolor{MyC}{C}}ases can be related to the \textbf{\textcolor{MyT}{T}}esting. In this example, the same test case is used. Observing the identical output of the print statements and the previously worked \textbf{\textcolor{MyC}{C}}ases allows the students to relate the \textbf{\textcolor{MyT}{T}}esting of their implementation to their previous \textbf{\textcolor{MyC}{C}}ases step. 

Two of the instructors highlighted students' reluctance to spend time planning their solutions on paper. One instructor wrote the following reflection:

\begin{quote}\footnotesize
    \emph{When it comes to the problem analysis, I see very few students trying to figure out the problem on paper first. They often try their luck in code and use a trial-and-error approach, basically trying to figure what could be the code that will satisfy the given test cases. This is not the right approach to coding, and I often have to force students to go back to analyzing the problem on paper.}
\end{quote}

Instead of \textbf{\textcolor{MyC}{C}}ases, one of the instructors found the \textbf{\textcolor{MyD}{D}}esign of Algorithm and \textbf{\textcolor{MyI}{I}}mplementation stages to be the main difficulty. He also highlighted that students were unwilling to spend time to plan their solutions before actually implementing them. This agrees with students' comments in the sense that students found the framework to be time consuming. But it is exactly this time spent in planning the solutions that enables students to successfully solve the problem. As one of the instructors put it:

\begin{quote}\footnotesize
    \emph{I genuinely think that the framework is not only helpful for novice programmers but for any programmer trying to solve a complex task. We actually got a lot of positive feedback for it. However, it is only useful for students who want to use it! The main difficulty consists of convincing the students to work on paper first. Moreover, many of them think it takes too much time doing so. }
\end{quote}

The instructors noticed a few pitfalls. In particular, the framework is demonstrated over an example. From a particular case, the instructor explains how it helps in deriving a rough draft of a pseudo-algorithm. For example, to explain how to write a Python program calculating the product of two matrices, after the problem statement, the instructor chooses two simple matrices, multiplies them on the board while orally decomposing the work process. This example helps in drafting a pseudo-algorithm and defining a first test case. However, some students tend to believe that testing this particular case is enough. Instructors should facilitate from the particular case to more general steps by discussing different \textbf{\textcolor{MyC}{C}}ases. Moreover, instructors must explicitly highlight the differences between the `C' and the `T'. The students should be reminded that the final phase is not only performed to catch the syntactic bugs of their programs or check whether the inputs corresponding to the particular case lead to the expected result. It is also a step to test the limits of their programs and see if it works with a more general case beyond what they did in the \textbf{\textcolor{MyC}{C}}ases step. For example, what happens if a list of lists (a typical matrix representation in Python) contains an empty sub-list? What if the values are not all numerical? What if the sub-lists do not have the same length? Encouraging the students to be `malicious' with their own functions seemed to have an impact on some students who then tried to build `unhackable' programs.

Moreover, it was recognised that the framework does not reduce the need to know the programming language and its syntax. In fact, an instructor stated that some students still struggled with the \textbf{\textcolor{MyD}{D}}esign of Algorithm and \textbf{\textcolor{MyI}{I}}mplementation stages. The reason for this difficulty is that many students are not familiar with the syntax. Therefore, as important as the problem solving steps are, the exercises and training on the language itself remain necessary for the whole PCDIT framework to be applied. This agrees with Linn and Clancey~\cite{Linn1992} who argued that both competencies are necessary.

\substepseparator

\noindent\textbf{Open Questions.} We are encouraged by the positive results in the survey, and the reflections from our students and instructors, which increase our confidence in the effectiveness of using the PCDIT problem solving framework in teaching. These results and reflections have prompted us to ask a number of further questions, many of which could be addressed in formal studies.

For example: must the framework be facilitated by a human instructor? Can the framework be used by students independently without any help from any instructor or can it only be used with some facilitation? Can the role of the instructor be replaced through other means in order to facilitate students' use of the framework? For example, can a more guided and detailed worksheet be used in this case? Can such facilitation be automated?

Another question of interest is regarding how we should implement and integrate this framework in our lessons? One of the highlighted issues is the perception that the PCDIT framework consumes a lot of time. How should the framework be implemented with such constraints and perceptions? Can we implement the framework in such a way that both students and instructors find the time taken is well spent? Would incorporating it into an AAT address this? How early should the framework be taught?

Lastly, we found that PCDIT might be useful not only for individual approaches to problem solving but also for group discussions when approaching a programming assignment. Further study should be done on how such framework can be used in a group setting and whether modifications are needed in the framework when collaborative learning is in place.

\section{Conclusion}
\label{sec:conclusions}

In this experience report, we presented PCDIT, a problem solving framework for novice programmers, designed to incorporate recommendations from the literature on improving metacognitive awareness. In particular, it encourages students to design/solve concrete cases well before any programming, and promotes regular reflections across the five main phases of the the process. We described our implementation of it for an introductory programming course, and showed how the framework's \textbf{\textcolor{MyC}{C}}ases step and non-linearity help novices to increase their confidence and better manage their cognitive load. In a post-use student survey, 62\% agreed or strongly agreed that it helped them to solve programming problems, with several highlighting that PCDIT helped them to organise their thoughts, solve problems systematically, and avoid jumping to conclusions without thinking through the problem first. Our instructors also reflected positively, highlighting a number of benefits, including PCDIT's clearly named steps (helping students reflect on where they are in the problem solving process), and the fact that its scaffolding explicitly mirrors the problem solving process we naturally follow (but that many novices don't). In future work, we plan to explore the use of the framework in the context of collaborative learning, and how its steps can be integrated into our automated assessments tools (similar to~\cite{Prather-et_al19a}).





\bibliographystyle{ieeetr}
\bibliography{new_ref}

\end{document}